\documentclass{JHEP3}
\usepackage{graphicx}
\usepackage{epsfig,amsmath,amsthm}
\newcommand{\be}{\begin{equation}}
\newcommand{\ee}{\end{equation}}
\newcommand{\bea}{\begin{eqnarray}}
\newcommand{\eea}{\end{eqnarray}}
\def\bse{\begin{subequations}}
\def\ese{\end{subequations}}

\def\IZ{\relax\ifmmode\hbox{Z\kern-.4em Z}\else{Z\kern-.4em Z}\fi}

\newcommand{\non}{\nonumber \\}

\def\half{\frac{1}{2}} \def\quart{\frac{1}{4}}

\def\del{{\partial}}

\def\cL{{\cal L}} 

  \def\eps{\epsilon}

\def\bi{\begin{itemize}} \def\ei{\end{itemize}}

\def\({\left(} \def\){\right)}
\def\[{\left[} \def\]{\right]}

\title{ \center{Non-Relativistic Gravitation: From Newton to Einstein and Back}}

\author{Barak Kol and Michael Smolkin\\
Racah Institute of Physics, Hebrew University\\
Jerusalem 91904, Israel\\
E-mail:
{\tt\href{mailto:barak_kol@phys.huji.ac.il}{barak\_kol@phys.huji.ac.il}},
\email{smolkinm@phys.huji.ac.il}}

\abstract{We present an improvement to the Classical Effective
Theory approach to the non-relativistic or Post-Newtonian
approximation of General Relativity. The ``potential metric
field'' is decomposed through a temporal Kaluza-Klein ansatz into
three NRG-fields: a scalar identified with the Newtonian
potential, a 3-vector corresponding to the gravito-magnetic vector
potential and a 3-tensor. The derivation of the
Einstein-Infeld-Hoffmann Lagrangian simplifies such that each term
corresponds to a single Feynman diagram providing a clear physical
interpretation. Spin interactions are 
 dominated by the exchange of the
gravito-magnetic field. Leading correction diagrams corresponding
to the 3PN correction to the spin-spin interaction and the 2.5PN
correction to the spin-orbit interaction are presented.}

\begin{document}

\section{Introduction}

In 2004 Rothstein and Goldberger \cite{GoldbergerRothstein1}
suggested a novel approach to describe gravity (within Einstein's
theory) for extended objects. Their method uses effective field
theory and replaces the extended object by a point particle, whose
interactions (or effective world-line action) encode all of its
physical properties, such as reaction to tidal gravitational
forces, ordered by a certain natural order of relevancy. This
method has the advantage of applying to General Relativity (GR)
tools which are normally associated with Quantum Field Theories
including Feynman diagrams, dimensional regularization and
effective actions. In particular action methods are more efficient
than studying the equations of motion.

Later that method was applied \cite{CGR} to determine the
thermodynamics of ``caged black holes'', namely small black holes
in the presence of compact extra dimensions. Caged black holes
were previously analyzed by means of Matched Asymptotic Expansion
(MAE) \cite{H4,dialogue1,dialogue2,KSSW1} due to their part in the
black-hole black-string phase transition
\cite{TopChange,review,HOrev} associated with the black string
instability of Gregory-Laflamme \cite{GL1}. The effective
field theory approach, while formally identical to MAE, typically
economizes the computation significantly. Recently an improved
 Classical Effective Field Theory (ClEFT) approach to
caged black hole appeared \cite{ClEFT-caged}.

The original main application of \cite{GoldbergerRothstein1} was
the non-relativistic motion of a binary system also known as the
Post-Newtonian (PN) approximation. In this paper
 we apply the improvements of \cite{ClEFT-caged} to this case, and especially
the stationary decomposition of fields. In section
\ref{NtoE-section} we start by recalling the evolution from
Newtonian gravity to Einstein's. In section \ref{NRG-section} we
consider Non-Relativistic Gravity (NRG), expanding Einstein's
theory as Newton's plus corrections.\footnote{The limit $v \ll c$
which is \emph{non-relativistic} from the point of view of
Einstein's fully relativistic theory, corresponds to
\emph{relativistic} corrections from the post-Newtonian
perspective, and it is also known by this latter name in the
literature.} We describe the proposal of
\cite{GoldbergerRothstein1} for an effective field theory of NRG
and we proceed to suggest an improvement via transforming to
NRG-fields. Finally in section \ref{app-section} we discuss two
applications: to the first post-Newtonian correction of the two
body problem, known as the Einstein-Infeld-Hoffmann (EIH)
Lagrangian, and to spin interactions.

{\bf Note added (v2)}. This published version includes several relatively
minor improvements. We detail the value of each individual
diagrammatic contribution to the EIH action (\ref{EIH}). We added
details about the diagrams for the spin interactions. Signs were
changed in the spin vertices (\ref{LOspin},\ref{NLOspin}) to
reflect current conventions. Finally, references were added and updated.

{\bf Version 3}. The generalization of the EIH Lagrangian to an
arbitrary dimension was computed recently in \cite{CDF} using the
EFT approach introduced in \cite{GoldbergerRothstein1}. We add a
subsection (\ref{higher_dim}) where the computation is done
through the use of our improved version of ClEFT. Our result
confirms all terms except for one which is to be corrected in a
revision of \cite{CDF}.

\section{From Newton to Einstein}
\label{NtoE-section}

Consider the motion of several masses whose sole interaction is
through gravity. Without loss of generality we will write the
actions for two masses. The Newtonian equations of motion
\cite{Principia} are concisely encoded by the action \be
 S=\int dt \[ \sum_{a=1}^2 \frac{m_a}{2}\, \dot{\vec{r}}_a^{~2} + \frac{G\, m_1\, m_2}{r}
 \] \label{Npart} \ee
 where $\vec{r}_1, \vec{r}_2$ are the locations of the two masses,
 and $r:=|\vec{r}|:=|\vec{r}_1- \vec{r}_2|$.

Introducing the gravitational field $\phi$, the familiar equations
of motion are encoded by an action which replaces the direct
gravitational potential in (\ref{Npart}) by a coupling of the
masses to $\phi$, together with a kinetic term for $\phi$ \be
 S = \int dt  \sum_{a=1}^2 \[ \frac{m_a}{2}\, \dot{\vec{r}}_a^{~2}
 - m_a\, \phi(\vec{r}_a) \] -\frac{1}{8 \pi G} \int dt\, d^3 x\, \(\vec{\nabla}
 \phi\)^2 \label{Nfield} ~. \ee

In Einstein's theory of gravity \cite{GR} the gravitational field
is promoted to a space-time metric $g_{\mu\nu}$. The two body
dynamics is given by the Einstein-Hilbert action together with the
relativistic action of point particles \be
 S=-\frac{1}{16 \pi G} \int \sqrt{-g}\, d^4x\, R[g] - \sum_{a=1}^2\, m_a\, \int d\tau_a ~.
  \label{EH} \ee
 where $x^\mu(\tau)$ is the particle's trajectory, the proper time is defined by
$\tau^2:=g_{\mu\nu}\, dx^\mu\, dx^\nu$, and for clarity we used
$c=1$ units. Here the two body problem becomes a 4d field dynamics
which is fully non-linear, and no closed solution is known or
believed to exist.

\section{Non-Relativistic Gravity: Stationary decomposition}
\label{NRG-section}

Consider the two body problem in General Relativity (GR) in the
case where both velocities are small with respect to the speed of
light. This problem applies to a binary inspiral process at its
early stages, which is a conjectured source for the widely sought
gravitational waves. The more traditional approach to the limit
within GR is known as ``the Post-Newtonian (PN) approximation''.

In 2004 Goldberger and Rothstein \cite{GoldbergerRothstein1} (see
also \cite{GoldbergerRothstein2} and a pedagogical introduction in
\cite{Goldberger-Lect}) pioneered an Effective Field Theory
approach to this problem. Denote a typical separation between the
masses by $r$, and a typical (small) velocity by $v$. Accordingly
the typical variation time for all fields is $r/v$. In space,
however, there are two typical lengths: $r$ and $r/v$. Accordingly
\cite{GoldbergerRothstein1} decomposes the metric into \be
 g_{\mu\nu} = H_{\mu\nu} + \bar{g}_{\mu\nu} \ee
 where $H_{\mu\nu}$ has length scale $r$ and since the time
variation scale is $r/v$, $H_{\mu\nu}$ is off-shell and it is
called the potential component, while $\bar{g}_{\mu\nu}$ is of
length scale $r/v$, it is on-shell and represents the radiation
component. This approach is referred to as ``Non-Relativistic GR''
(NRGR) and can also be called ``Non-Relativistic Gravity'' (NRG).

In this paper we concentrate on the potential component
$H_{\mu\nu}$. According to \cite{GoldbergerRothstein1} its
propagator includes only the spatial frequencies and not the
temporal frequencies, which are subleading in the non-relativistic
limit and hence treated as a perturbation. As $H$ is considered
static for the purposes of the propagator, it is natural to
transform the fields through performing a temporal Kaluza-Klein
dimensional reduction as in \cite{ClEFT-caged} \be
 ds^2 = e^{2 \phi}(dt - A_i\, dx^i)^2 -e^{-2 \phi}\, \gamma_{ij}\,
 dx^i dx^j ~. \label{KKansatz} \ee
This relation defines a change of variables from $g_{\mu\nu}$ to
$(\gamma_{ij},A_i,\phi), ~i,j=1,2,3$ which we call ``NRG-fields''.

The action, when translated into NRG-fields, and within the static
approximation, namely when all fields are $t$-independent becomes
\be
 S = -\frac{1}{16\pi G} \int dt\, dx^3 \sqrt{\gamma}
  \[ R[\gamma] + 2\, \(\del \phi\)^2 -  \frac{1}{4}\, e^{4\phi} F^2 \]~,\label{dim-red-action} \ee
 where $\(\del \phi\)^2 = \gamma^{ij}\, \del_i \phi\,
\del_j \phi=(\vec{\nabla} \phi)^2+\dots$ and the term with the
 field strength $F$ is conventionally defined by $F^2=F_{ij}
F^{ij}, ~~F_{ij}=\del_i A_j- \del_j A_i$.

Let us discuss the physical meaning of the new fields $\phi,
A_i,~\gamma_{ij}$. Comparing the kinetic term for $\phi$ in the
action (\ref{dim-red-action}), with the Newtonian field action
(\ref{Nfield}) we find it natural to identify $\phi$ with the
Newtonian potential (this is obvious when the 3-metric is flat
$\gamma_{ij}=\delta_{ij})$. In this sense we are back to Newton
(plus correction terms). The constant pre-factor $2$ in this
kinetic term is related to the polarization dependence of the $g$
propagator (the original graviton): if we were to compute the
Newtonian potential in the original action, prior to dimensional
reduction, this same factor would have emerged from the graviton
propagator in the standard Feynman gauge.

The vector potential $A_i$ has an action which resembles the
Maxwell action in 3d, and accordingly it is natural to call $F$
the gravito-magnetic field. This name originates in a certain
similarity between gravity and electro-magnetism. The strong
similarity between Newton's gravitational force and Coulomb's
static electrical force, together with the observation that the
transition from electro-statics to electro-dynamics requires to
supplement the scalar electric potential by a vector potential,
promoted already in the 19th century suggestions to add a vector
potential to the gravitational degrees of freedom. It is in fact
known how to obtain such a vector potential in the weak gravity/
Post-Newtonian approximation to GR, a point of view known as
``Gravito-Electro-Magnetism (GEM)'' (see \cite{wikiGEM} and
references within). We note the reversed sign of the kinetic term
for $F$. This is directly related to the fact that the spin-spin
force in gravity has an opposite sign relative to
electro-dynamics, namely ``north poles attract'' \cite{Wald-spin}.

Finally the 3-metric tensor $\gamma_{ij}$ comes with a standard
Einstein-Hilbert action in 3d (this is achieved through the Weyl
rescaling factor in front of $\gamma_{ij}$ in the ansatz).

Once time-dependence is permitted, the action contains time
derivatives. Of particular interest are terms quadratic in the
fields (including time derivatives) which are considered as
vertices rather than being part of the propagator as mentioned
above. The only such term which will be required here is \be
 S \supset \frac{1}{16\pi G} \int dt\, dx^3\, 2 \dot{\phi}^2 \label{phi-dot} \ee
 rendering the $\phi^2$ sector Lorentz invariant. It is obtained after decoupling $\phi$ and $A$ at the quadratic
level through the use of a Lorentz invariant gauge-fixing term.
 In addition we note that the definition of $\phi, A_i,\,
\gamma_{ij}$ (\ref{KKansatz}) is given here in the $t$-independent
case, and when time dependence is incorporated it is conceivable
that it should be supplemented by terms with time derivatives.


In the new variables (\ref{KKansatz}) the point-particle action
which appears in  (\ref{EH}) becomes \bea
 S_{pp} \equiv -m_0 \int d\tau &=& -m_0 \int dt\, e^\phi\,
 \sqrt{(1-\vec{A} \cdot \vec{v})^2-e^{-4\phi}\, \gamma_{ij}\,
 v^i\, v^j } = \non
 &=& -m_0 \int dt\, \(1 -\half\, v^2+ \phi - \vec{A} \cdot \vec{v} + \frac{3}{2}\, \phi\, v^2 + \dots \) \label{pp}
 \eea
 where $\vec{v} \equiv \dot{\vec{r}}$ is the velocity vector. The change of variables
has the advantage that the propagator is diagonal with respect to
the field $\phi$ which couples to the world-line at lowest order
(through the interaction $(-)m_0\, \phi$ ).

For a spinning object the lowest order interaction, found in
\cite{Porto1} and translated to NRG-fields in \cite{ClEFT-caged}
reads (in the conventions of \cite{ClEFT-caged}) \be
 S \supset \frac{1}{4} \int dt\, J_0^{ij}\, F_{ij} = \half\, \vec{J}
 \cdot \vec{B} \label{LOspin} \ee
where we denoted the angular momentum vector $J_i=\eps_{ijk}
J^{jk}/2$ and the gravito-magnetic field strength $B_i=\eps_{ijk}
F^{jk}/2$. Additional terms come with additional fields \be
 S \supset \int dt\, J_0^{ij} \( F_{ij}\, \phi - \half\, A_i\, \del_j \phi - \quart \delta\gamma_j^{~k}\, F_{ik} \)
 \label{NLOspin}
\ee Yet other terms come with powers of $v$ (by departing from
stationarity).

\section{Applications}
\label{app-section}

Here we apply the new NRG-fields to obtain valuable insight into
the Einstein-Infeld-Hoffmann Lagrangian and into spin interactions
within NRG.

\subsection{Einstein-Infeld-Hoffmann}

When passing from electro-statics to electro-dynamics we may
integrate out the electro-magnetic field and obtain an action for
two interacting charged particles which depends on their
velocities as well as their locations thus including the effects
of both electric and magnetic forces. The analogous 1PN correction
to the Newtonian gravitational action (\ref{Npart}) is called
Einstein-Infeld-Hoffmann (EIH) \cite{EIH} \be
 \cL_{EIH} = \frac{1}{8} \sum_{a=1}^2 m_a\, \vec{v}_a^4 + \frac{G
 m_1 m_2}{2 r} \[3 (\vec{v_1}^2+  \vec{v_2}^2)-8 \vec{v_1} \cdot
 \vec{v_2} + \vec{v_1}_\perp \cdot \vec{v_2}_\perp  \]
 -\frac{G^2\, m_1 m_2 (m_1+m_2)}{2\, r^2} \label{EIH} \ee
 where $\vec{v_i}_\perp := \vec{v_i}-\vec{r}\, (\vec{v_i}\cdot \vec{r})/r^2$.
From an intuitive Newtonian perspective, this correction
represents the contribution to the gravitational interaction from
both kinetic and potential energies, as well as a correction
accounting for the finite speed of light.

\begin{figure} [t!] \centering \noindent
\includegraphics[width=6cm] {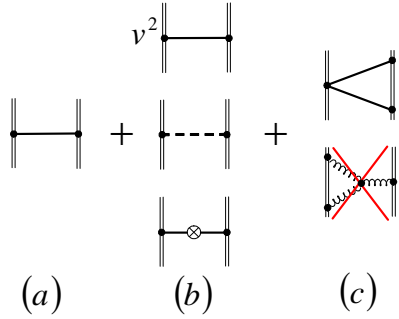}
\caption[]{Feynman diagrams contributing to the Newtonian two body
Lagrangian (\ref{Npart}) and the 1PN Einstein-Infeld-Hoffmann
Lagrangian (\ref{EIH}). Double lines represent the masses and do
not carry propagators. Solid lines represent $\phi$, the Newtonian
potential, dashed lines represent $A_i$, the vector potential,
while spring-like lines are non-discriminating notation for all
the polarizations of the original graviton $g$. A power of $v$
near a vertex denotes that the vertex was expanded in $v$, while a
circled cross denoted $v$-dependent correction to the (static)
propagator. A dot vertex for a dashed line represent the $\vec{v}
\cdot \vec{A}$ vertex of the world-line action (\ref{pp}). Diagram
(a) is the Newtonian potential. Each diagram in (b,c) matches a
specific terms in the EIH Lagrangian (\ref{EIH}) as described in
the text. Diagrams in (b) represent the $v$ dependent EIH terms,
while $(c)$ are $v$-independent. Note that the relatively
complicated triple vertex diagram, fig. 5(a) of
\cite{GoldbergerRothstein1}, disappeared.} \label{EIH-fig}
\end{figure}

In \cite{GoldbergerRothstein1} the calculation of this action,
whose associated equations of motion required some 36 pages of
traditional GR methods in 1938, was reduced to the computation of
5 Feynman diagrams over less than 2 pages.\footnote{Obviously quite a number of works
touched upon the EIH result since 1938 and we shall not attempt to provide a comprehensive list. At the suggestion
of a referee we would like to mention the ``ADM Hamiltonian approach'' \cite{ADM-Ham} which yielded the two body Hamiltonian up to 2PN through ``tedious calculations''. For field theoretic approaches which predate \cite{GoldbergerRothstein1} see references therein.}
 After translation into
NRG-fields we obtain the Feynman diagrams shown in figure
\ref{EIH-fig}. The first pay-off is that the triple vertex diagram
(5(a) of \cite{GoldbergerRothstein1}) gets eliminated, since there
is no cubic vertex for $\phi$ in our action
(\ref{dim-red-action}). Noting that both figures 5(a) and 5(b) of
\cite{GoldbergerRothstein1} are proportional to the last term in
the EIH action (\ref{EIH}), it is not surprising that they can be
replaced by the single diagram in fig. \ref{EIH-fig}(c). This is
especially fortunate since the eliminated diagram was the only one
to use the awkward 3-graviton vertex and the only one including a
loop. As a result of this economization, each diagram in figure
\ref{EIH-fig} is responsible for precisely one term in
(\ref{EIH}): the $v^4$ term comes from the kinetic part of
(\ref{pp}); the next 3 terms all proportional to $v^2$ come from
the diagrams in figure \ref{EIH-fig}(b): the $v_1^2+v_2^2$ term
comes from the top diagram, the $\vec{v_1} \cdot \vec{v_2}$ term
comes from the middle diagram and the $\vec{v_1}_\perp \cdot
\vec{v_2}_\perp$ comes from the bottom diagram where the vertex
(\ref{phi-dot}) is used; finally the last term of (\ref{EIH})
comes from the top diagram of figure \ref{EIH-fig}(c).

Additional pay-off comes in terms of insight into the fields which
propagate in the diagram. Almost all of them are the gravitational
potential $\phi$. An exception is the second from top diagram in
fig. \ref{EIH-fig}(b) , where the vector potential propagates and
it is responsible for the $v_1\, v_2$ factor coming from both
vertices.

\subsubsection {In higher dimensions}
\label{higher_dim}

Following the recent generalization of the EIH Lagrangian to an
arbitrary space-time dimension $d$ \cite{CDF} using the EFT
approach introduced in \cite{GoldbergerRothstein1}, we perform the
computation within the improved version of ClEFT discussed above,
and compare the results.

The $d$-dimensional analogs of (\ref{KKansatz}, \ref{dim-red-action},
\ref{phi-dot}) and
(\ref{pp}) are \be
 ds^2 = e^{2 \phi}(dt - A_i\, dx^i)^2 - e^{-2\phi / (d-3)}\, \gamma_{ij}\,
 dx^i dx^j ~, \label{KKansatzin-in-D} \ee
 \be
 S = -\frac{1}{16\pi G} \int dt\, d^{d-1}x \sqrt{\gamma}
  \[ R[\gamma] + {d-2 \over d-3} \, \(\del \phi\)^2 -  \frac{1}{4}\, e^{2(d-2)\phi/(d-3)} F^2 \]
  \label{dim-red-action-in-D} \ee
\be
 S \supset \frac{1}{16\pi G} \int dt\, d^{d-1}x\, {d-2 \over d-3}\, \dot{\phi}^2~. \label{phi-dot-in-D} \ee
and \bea
 S_{pp} \equiv -m_0 \int d\tau &=& -m_0 \int dt\, e^\phi\,
 \sqrt{(1-\vec{A} \cdot \vec{v})^2-e^{-2(d-2)\phi/(d-3)}\, \gamma_{ij}\,
 v^i\, v^j } = \non
 &=& -m_0 \int dt\, \(1 -\half\, v^2+ \phi - \vec{A} \cdot \vec{v} + \frac{d-1}{2(d-3)}\, \phi\, v^2 + \dots \) \label{pp-in-D}
 \eea

The Feynman diagrams contributing to the 1PN correction to the
Newtonian gravitational action are the same diagrams as in 4d, namely those shown in figure \ref{EIH-fig}.
The triple vertex diagram gets eliminated as previously. As a result
the $d$-dimensional generalization of EIH Lagrangian is given by
\bea
 \cL_{EIH} &=& \frac{1}{8} \sum_{a=1}^2 m_a\, \vec{v}_a^{\,4} \non
 &+& \frac{1}{2} U_{N}(r)
\[\frac{d-1 }{(d-3)} (\vec{v_1}^2+  \vec{v_2}^2)-\frac{4(d-2)}{
d-3} \vec{v_1} \cdot
 \vec{v_2} +
 \( \vec{v_1} \cdot \vec{v_2}- (d-3) (\vec{v_1} \cdot \hat r) (\vec{v_2} \cdot \hat
 r) \) \]   \non 
 &-& \frac{m_1+m_2}{2 m_1 m_2}\, U_N(r)^2 \label{EIH-in-D}
 \eea
where $U_{N}(r)$ is the Newtonian potential energy corresponding to
figure \ref{EIH-fig}(a)
 \be U_{N}(r)= 2\, {d-3 \over d-2}\, \frac{ \Gamma(\frac{d-3}{2}) }{ \pi^{(d-3)/2} }\, \frac{G
 m_1 m_2}{r^{d-3}} = \frac{8 \pi}{(d-2) \Omega_{d-2}}\, \frac{G
 m_1 m_2}{r^{d-3}}~, \ee
 and $\Omega_{d-2}$ denotes the volume of the $d-2$ dimensional sphere.
The $v^4$ term originates from the correction to the kinetic
energy and comes from the expansion of (\ref{pp-in-D}); the terms
on second line are of order $v^2$ and come from the diagrams in
figure \ref{EIH-fig}(b): the first term  proportional to
$v_1^2+v_2^2$ comes from the top diagram, the second term
proportional to $\vec{v_1} \cdot \vec{v_2}$ comes from the middle
diagram while the last terms come from the bottom diagram; finally
the term on the third line comes from the top diagram of figure
figure \ref{EIH-fig}(c).

A comparison with \cite{CDF} reveals that our computation confirms
theirs for all diagrams but one: the numerical coefficient in the
term corresponding to figure \ref{EIH-fig}(c) is different. We are
told that it will be corrected in a revision of \cite{CDF}.


\subsection{Spin interactions}

NRG-fields will significantly simplify the computation of spin
interactions, and stress the role of the gravito-magnetic field.
Here we limit ourselves to a discussion of the relevant diagrams
in terms of NRG-fields, leaving for the future the detailed
calculation which involves issues such as the spin supplementary
condition (the relation between $J_0^{ij}$ and $\vec{v}$) whose
current presentation in the literature leaves room for improved
understanding, in our opinion.



\begin{figure} [t!] \centering \noindent
\includegraphics[width=6cm] {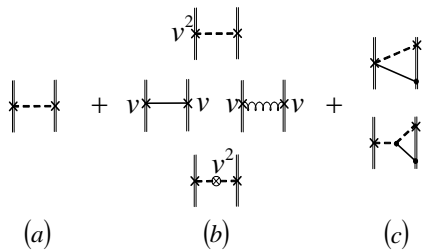}
\caption[]{Diagrams representing spin-spin interactions.  The
cross vertex represents the $\vec{J} \cdot \vec{B}$ vertex
(\ref{LOspin}). Unlike the previous figure, here the spring-like
line represents the 3-tensor $\gamma$. Other notation remains the
same. Diagram (a) represents the leading order at 2PN, in terms of
the gravito-magnetic field. The other diagrams represent the next
to leading contributions at 3PN: (b) represent $v^2$ corrections,
while $(c)$ represent $Gm/r$ corrections.} \label{fig2}
\end{figure}

The leading spin-spin interaction \be
 S_{ss} = G \int dt\, \frac{\vec{J_1} \cdot \vec{J_2} - 3 (\vec{J_1} \cdot \hat{r})
 (\vec{J_2} \cdot \hat{r})}{r^3} \ee where $\hat{r}:=\vec{r}/r$,
is given by the Feynman diagram in figure \ref{fig2}(a) and is
seen to consist of an exchange of the gravito-magnetic field $A_i$
at order 2PN.

\begin{figure} [t!] \centering \noindent
\includegraphics[width=6cm] {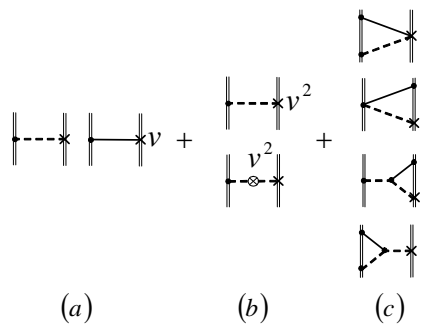}
\caption[]{Diagrams representing contributions to the spin-orbit
interaction. Diagrams (a) represent the leading order at 1.5PN.
The diagrams in (b) and (c) are a sample of those representing the
first correction at 2.5PN. The notation is the same as in the
previous figures.} \label{fig3}
\end{figure}

The next to leading spin-spin interaction was computed in
\cite{PortoRothstein} (see also
\cite{SteinhoffHergtSchafer,PRcomment,PortoRothstein2}) in terms
of 5 diagrams (fig. 1 and fig.2 of \cite{PortoRothstein}) which
are quite analogous to the 5 diagrams for EIH in
\cite{GoldbergerRothstein1}. In terms of NRG-fields we obtain the
diagrams of figure \ref{fig3}. The pay-off is that the
``voluminous'' 3-graviton vertex in fig. 2(a) of
\cite{PortoRothstein}  is replaced by the compact $\phi\, F^2$
cubic vertex which can be read from the dimensionally reduced
action (\ref{dim-red-action}) avoiding the need for a symbolic
manipulation program. Actually a similar diagram appeared in the
calculation of the vanishing renormalization of the angular
momentum in \cite{ClEFT-caged}. Following the  appearance of the
first arXiv version of this paper, the next to leading spin-spin
interaction was computed with NRG-fields \cite{Levi} and was found
to considerably simplify the calculations.

The situation for the spin-orbit interaction seems to be quite
similar. The leading contribution is given now by the two diagrams
in figure \ref{fig3}(a), which are related through Galilei
invariance. The next to leading terms at order 2.5PN were computed
in \cite{SO-NLO} in terms of equations of motion and in \cite{DJS}
in terms of a Hamiltonian. Here we limit ourselves to pointing out
some of the diagrams which would appear at this order in figures
\ref{fig3}(b,c).


\subsection*{Acknowledgements}

This research is supported by The Israel Science Foundation grant
no 607/05,  DIP grant H.52, EU grant MRTN-CT-2004-512194 and the
Einstein Center at the Hebrew University.

\end{document}